# Agile Test-based Modeling

Bernhard Rumpe
Software Systems Engineering
TU Braunschweig, Germany
www.sse.cs.tu-bs.de

*Model driven architecture (MDA) concentrates on the use of models during software development. An approach using models as the central development artifact is more abstract, more compact and thus more effective and probably also less error prone. Although the ideas of MDA exist already for years, there is still much to improve in the development process as well as the underlying techniques and tools. Therefore, this paper is a follow up on [13], reexamining und updating the statements made there. Here two major and strongly related techniques are identified and discussed: Test case modeling and an evolutionary approach to model transformation.*

## 1 Introduction: Modeling meets Programming

The UML [1] has become the most popular modeling language for software intensive systems. UML as well as other models (Matlab or domain specific ones) can be used for quite a variety of purposes. Among them diagrams are still mainly used for *documentation* of requirements and design. Requirements are usually captured in natural language and a few informal and top-level drawings that denote an abstract architecture, use cases or activity diagrams. Architecture and designs are then captured and documented with models. In practice, these models are increasingly often used for *generation* of code respectively code frames that are filled in manually.

More sophisticated and therefore less widespread uses of models are analysis of certain features (such as throughput, robustness, failure likelihood), generation of tests from models and a transformation based evolutionary approach from high-level models to running code. Quite a few UML-based tools offer functionality to emulate models or generate code or at least code frames. Tool vendors still work hard on continuous improvement of these features. It is foreseeable that together with the effort of defining virtual machines respectively executable UML, a large sublanguage of the UML will become a high-level programming language and modeling at this level becomes identical to programming. This raises a number of interesting questions:

- Is it critical for a modeling language to be also used as programming language? For example analysis and design models may become overloaded with details that are not of interest yet, because modelers are addicted to executability. To our experience, high-level designs then become too detailed to be reused in different applications. As a consequence it would be necessary to integrate an executable sublanguage of the UML with a specification-oriented and not-executable language for reuse of requirement captures.
- Is the UML expressive enough to describe systems completely or will it be accompanied by conventional languages? How well are these integrated? We believe that there will be techniques to thoroughly integrate several languages. Today, executable pieces of code written e.g. in Java are still treated like strings within UML models. In the future, there will be better integration and specifically, there will be techniques for composition of such language parts.
- How will the toolset of the future look like and how will it overcome round trip engineering (i.e. mapping code and diagrams in both directions)? We believe that round-trip engineering is not a technique that will last for ever. Round-trip had been used in a similar way, when people didn't trust compilers and wanted to check and change compiled code manually. But today, it's quite unclear to understand how to overcome this kind of jojo-engineering.
- What are the implications of an executable UML on the development process? Only few people believe in the UML as a full programming language. Indeed, the UML (including its action language, OCL and behavioral diagrams) is in its current form not very convenient for programming. However, if tooling becomes more efficient and the UML is enriched with appropriate programming concepts, this might change. But will it lead to better code or just more quickly to bad implementations?

In [4,5] we have discussed these issues and have demonstrated, how the UML in combination with Java may be used as a high-level programming language. But, to ensure quality of the result the UML cannot only be used for modeling the application, but more importantly for modeling tests on various levels (class, integration, and system tests) as well.

One advantage of using models for test case description is that application specific parts are modeled with UML-diagrams and technical issues, such as connection to frameworks, error handling, persistence, or communication can be handled by the parameterized code generator. This basically allows us to develop models independent of any technology or platform, as for example proposed in [6]. Only in the generation process platform dependent elements are added. When the technology changes, we only need to update the generator, but the application defining models can be reused without much change. This concept also supports the above mentioned MDA-Approach [2] of the OMG. Another important advantage is that both, the production code and automatically executable tests at any level, are modeled by the same kind of UML diagrams. Therefore developers use a single homogeneous language to describe implementation and tests. This will enhance the availability of tests already at the beginning of the coding activities. Similar to the "test first approach" [7,8], sequence diagrams are used to model drives of test cases. They can be taken from requirements that have also been modeled using sequence diagrams.

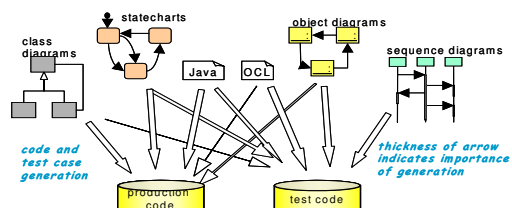

**Fig. 1.** Mapping of UML-models to code and test code Some of the UML-models (mainly deployment and class diagrams as well as statecharts) are used constructively, while others are used for test case definition (mainly OCL, sequence and enhanced object diagrams). The following Fig. 1 illustrates the key mappings from various diagrams to the production and test code. The notations on the left side are usually considered "complete" and are therefore useful for generating production code, whereas the notations on the right are "exemplaric" and thus useful for modeling individual tests. However, both sides are not disjoint, because it's possible to generate e.g. initialization code from an object diagram as well as test drivers from a statechart.

The following section 2 discusses the combination of agile methods and model-based software development from a methodical point of view. It is argued that the use of models increases efficiency, quality and other project elements such that the project can be downsized and run in an agile way. As primary technical elements of model-based development, the definition of model-based tests is discussed in section 3 and the evolution (refactoring) of models is discussed in section 4. Section 5 gives a final conclusion and summary.

## 2 Agile Modeling: Using Models in Agile Projects

In the last years a number of Agile Methods [9] have been brought to practice that share a some special characteristics subsumed under "agile" resp. "light-weight". Among these Extreme Programming (XP) [3] is currently the most widely used and discussed method [17]. Some of the XP characteristics are:

- XP early focuses on the primary goal, the running production code. Other artifacts, like documentation are produced and used only in a very limited way, but coding standards are enforced to document the code.
- At any stage of development, automated tests are used to ensure quality of the result. Our practical experience shows, that when this is properly done, the defect rate is in fact considerably low. Furthermore, the automation allows us as well as new developers and the paying customer to repeat tests continuously, even if the customer doesn't understand the content of the test.
- Very small iterations with continuous integration are enforced and the system is kept as simple as possible.
- Refactoring of code is used to improve the code structure and tests ensure the defect rate introduced through refactoring is rather small if existent at all.

The lack of documentation is motivated by the reduction of workload gained and the observation, that developers don't trust documents anyway, because these are out of date too often. So, XP focuses on code. All design activities manifest in the code directly. Quality is ensured through strong emphasis on testing activities, ideally on development of the tests before the production code
("test first approach" [7]). An explicit architectural design phase is abandoned and the architecture emerges during coding. Architectural shortcomings are resolved through the application of refactoring techniques [10,11]. These are transformational techniques to refactor a system in small steps to enhance its structure. The concept isn't new [12], but through availability of appropriate tools and the use of refactoring in XP, transformational development now has the potential to become used in a wider range of projects.

When using an executable version of UML to develop the system within an agile approach, the development project should become even more efficient. On the one hand, through the abstraction of the platform independent models, these models are more compact. When developing such a model, the developer can focus on requirements only, completely disregarding the technological platform. These models can more easily be written, read and understood than code. On the other hand in classic development projects these models are developed for documentation anyway. But, increased reuse of these models for later stages now becomes feasible through better assistance. Therefore, model-based development as proposed by the MDA-approach [2] should become applicable in recent future. These UML-models also serve as up-to-date documentation much better than commented code does.

## 3 Model-based Testing

There exists quite a variety of testing strategies [14,15]. The use of models for the definition of tests and production code can be manifold:
- Code or at least code frames can be generated from a design model.
- Test cases can be derived from an analysis or design model that is not used/usable for constructive generation of production code. For example behavioral models, such as statecharts, can be used to derive test cases that cover states, transitions or even paths.
- The modeling technique itself can be used to describe a test case or at least a part thereof.

The first two forms have already been discussed e.g. in [15]. Therefore, in this section we concentrate on the development of models that describe tests. A typical test, as shown in Fig. 2 consists of a description of the test data, the test driver and an oracle characterizing the desired test result. In object-oriented environments, the test data can usually be described by an object diagram (OD). It shows the objects necessary to run the test as well as concrete values for their attributes and the linking structure. The test driver can be modeled using a simple method call or, if more complex, a sequence diagram (SD). An SD has the considerable advantage that not only the triggering method calls can be described, but it is possible to model desired interactions and check object states during the test run.

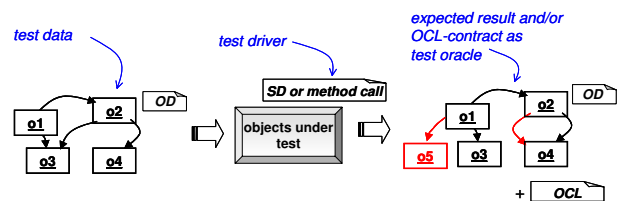

**Fig. 2.** Structure of a test modeled with object diagrams (OD), sequence diagram (SD) and the Object Constraint Language (OCL).

For this purpose, the Object Constraint Language (OCL, [16]) is used. Furthermore, it has proven efficient to model test oracles using a combination of an object diagram and OCL properties. The object diagram in this case serves as a property description and can therefore be rather incomplete, just focusing on the desired effects. The OCL constraints used can also be general invariants or specific property descriptions.

As already mentioned, being able to use the same, coherent language to model the production system and the tests allows for a good integration between both tasks. It allows the developer to immediately define tests for the constructive model developed. It is imaginable that in a kind of "test-first modeling approach" the test data in form of possible object structures is developed before the actual implementation.

# 4 Model Evolution using Automated Tests

Neither code nor models are correct from the beginning. For code, many sources of incorrectness can rather easily be analyzed using type checkers of compilers and automated tests that run on the code. For models this is usually a problem that leaves many errors undetected in analysis and design models. This is particularly critical as conceptual errors in these models are rather expensive if detected only late in the development process. The use of code generation and automated tests helps to identify errors in these models.

Besides detecting errors, which might even result from considerable architectural flaws, nowadays, it is expected that the development and maintenance process is capable of being flexible enough to dynamically react on changing requirements. In particular, enhanced business logic or additional functionality should be added rapidly to existing systems, without necessarily undergo a major re-development or re-engineering phase. This can be achieved at best, if techniques are available that systematically evolve the system using transformations. To make such an approach manageable, the refactoring techniques for Java [10] have proven that a comprehensible set of small and systematically applicable transformation rules seems optimal. Transformations, however, cannot only be applied to code, but to any kind of model. A number of possible applications are discussed in [12].

Having a comprehensible set of model transformations at hand, model evolution becomes a crucial step in software development and maintenance. Architectural and design flaws can then be more easily corrected, superfluous functionality and structure removed, structure for additional functionality or behavioral optimizations be adapted, because models are more abstract, exhibit higher-level architectural and design information in a better way.

Two simple transformation rules on a class diagram are shown in Fig. 3. The figure shows two steps that move a method and an attribute upward in the inheritance hierarchy. The upward move of the attribute is accompanied by the only context condition, that the other class "Guest" didn't have an attribute with the same name yet. In contrast, moving the method may be more involved. In particular, if both existing method bodies are different, there are several possibilities: (1) Move up one method implementation and have it overridden in the other class. (2) Just add the method as abstract signature in the superclass. (3) Adapt the method implementations in such a way that common parts can be moved upward. This can for example be achieved by factoring differences between the two implementations of "checkPasswd" into smaller methods, such that at the end a common method body for "checkPasswd" remains. As a context condition, the moved method may not use attributes that are available in the subclasses only.

Many of the necessary transformation steps are as simple as the upward move of an attribute. However, others are more involved and their application comes with a larger set of context conditions and accompanying steps similar to the adaptation necessary for the "checkPasswd" method. These of course need automated assistance. The power of these simple and manageable transformation steps comes from the possibility to combine them and evolve complex designs in a systematic and traceable way.

Following the definition on refactoring [10], we use transformational steps for structure enhancement that does not affect "externally visible behavior". For example both transformations shown in Fig. 3 do not affect the external behavior if made properly.

By "externally visible behavior" Fowler in [10] basically refers to behavioral changes visible to the user. This can be generalized by introducing an abstract "system border". This border serves as interface to the user, but may also act as interface to other systems. Furthermore, in a hierarchically structured system, we may enforce behavioral equivalence for "subsystem borders" already. It is therefore necessary to explicitly describe, which kind of behavior is regarded as externally visible. For this purpose tests are the appropriate technique to describe behavior, because (1) tests are already available through the development process and (2) tests are automated which allows us to check the effect of a transformation through inexpensive, automated regression testing. A test case thus acts as an "observer" of the behavior of a system under a certain condition. This condition is also described by the test case, namely through the setup, the test driver and the observations made by the test. Tests do not necessarily constrain their observation to "externally visible behavior", but can make observations on local structure, internal interactions or state properties even during the system run. Therefore, it is essential to identify, which tests are regarded as "internal" and are evolving together with the transformed system and which tests need to remain unchanged, because they describe external properties of the system. Tests in one categorization can roughly be divided into unit tests, integration tests and acceptance tests.

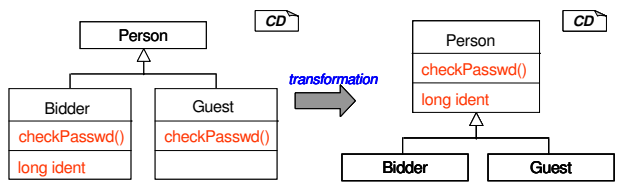

**Fig. 3.** Two transformational steps moving an attribute and a method along the hierarchy.

*Unit and integration tests* focus on small parts of the system (classes or subsystems) and usually take a deep look into system internals. It therefore isn't surprising that these kinds of tests can become erroneous after a transformation of the underlying models. Indeed, these tests are usually transformed together with the code models. For example, moving an attribute upward as shown in Fig. 3 induces object diagrams with Guest-objects to be adapted accordingly by providing a concrete value for that attribute. In this case it may even be of interest to clone tests in order to allow for different values to be tested. Contrary, tests may also become obsolete if functionality or data structure is simplified. The task of transforming test models together with production code models can therefore not be fully automated.

Unit and integration tests are usually provided by the developer or test teams that have access to the systems internal details. Therefore, these are usually "glass box tests". *Acceptance tests*, instead, are "black box" tests that are provided by the user (although again realized by developers) and describe external properties of the system. These tests must be a lot more robust against changes of internal structure.

To achieve robustness, acceptance tests should be modeled against the published interfaces of a system. In this context "published" means that parts of the system that are explicitly marked as externally visible and therefore usually rather stable. Only explicit changes of requirements lead to changes of these tests and indeed the adaptation of requirements can very well be demonstrated through adaptation of these test models followed by the transformations necessary to meet these tests afterwards in a "test-first-approach". An adapted approach also works for changes in the interfaces between subsystems.

## 5 Conclusions

The presented approach can be summarized as a pragmatic method to model-based software development. It suggests the use of models as primary artifact for requirements and design documentation, code generation and test case development. Transformations on models allow an efficient adaptation of the system to changing requirements and technology, optimizing architectural design and fixing bugs. To ensure the quality of such an evolving system, intensive sets of test cases are a must. They are modeled using the same language (UML) and thus exhibit a good integration and allow to model system and tests in parallel.

However, the methodology sketched here was in its basics already defined in [13], but still is a major proposal. Major efforts still need to be undertaken. The technology for transformation of models is not mature yet. Neither are the tools ready for major practical use, nor are semantically useful transformations understood in all their details. Neither the pragmatic methodology, nor the underpinning theory are very well explored yet.

To summarize, models can be used as described in this paper, but of course there are other possibilities of use. For example, it should be possible to have a variety of sophisticated analysis and manipulation techniques available that ideally operate on the same notations, but are used for requirements validation in early stages of the project.